# Dynamical Volatilities for Yen-Dollar Exchange Rates

Kyungsik Kim*, Seong-Min Yoon [a], C. Christopher Lee [b] and Myung-Kul Yum [c]

*Department of Physics, Pukyong National University, Pusan 608-737, Korea*

[a]*Division of Economics, Pukyong National University, Pusan 608-737, Korea*

[b]*Department of Business Administration, Central Washington University, WA 98926, USA*

[c]*Department of Pediatric Cardiology, Hanyang University, Kuri 471-701, Korea*

ABSTRACT

We study the continuous time random walk theory from financial tick data of the yen-dollar exchange rate transacted at the Japanese financial market. The dynamical behavior of returns and volatilities in this case is particularly treated at the long-time limit. We find that the volatility for prices shows a power-law with anomalous scaling exponent $\kappa$ = 0.96 (one minute) and 0.86 (ten minutes), and that our behavior occurs in the subdiffusive process. Our result presented will be compared with that of recent numerical calculations.

*Corresponding author. Tel: +82-51-620-6354; Fax: +82-51-611-6357.
E-mail address: kskim@pknu.ac.kr.

## 1. INTRODUCTION

The investigation of the continuous time random walk (CTRW) theory between economists and physicists has recently received considerable attention as one interdisciplinary field [1,2]. This subject has particularly led to a better understanding for novel universal properties based on statistical concepts and methods. Particularly, one of prominent problems for the regular and disordered systems is the random walk theory [3] in the stochastic process. This theory has extensively developed to the CTRW theory, formerly introduced by Montroll and Weiss [4], which is essentially characterized both by the transition probability dependent of the length between steps and by the distribution of the pausing times [5]. Until now, the CTRW theory has even more extended to study in natural, applied, and social sciences, and among many outstanding topics, there have been mainly concentrated on the reactive and strange kinetics [6,7], fractional diffusion equations [8], random networks, earthquake modeling, hydrology, and financial options [9].

Scalas [10] has recently presented the correlation function for bond walks from the time series of bond and BTP (Buoni del tesoro Poliennali) futures exchanged at the London International Financial Futures and Options Exchange (LIFFE). Scalas *et al*. [11] have discussed that CTRW theory is applied to the dynamical behavior for tick-by-tick data in financial markets. Mainardi *et al*. [12] have discussed the waiting-time distribution for bond futures traded at LIFFE. Kim and Yoon [13] investigated the dynamical behavior of volume tick data for the bond futures in the Korean Futures Exchange market and obtained that the decay distribution for the survival probability is exhibited novel stretched-exponential forms. Scalas *et al*. [14] applied the CTRW theory to models of the high-frequency price dynamics. They reported that the survival probability for high-frequency data of 30DJIA stocks shows an non-exponential decay.

Our purpose of this paper is to present theoretical and numerical arguments for the yen-dollar exchange rate traded at the Japanese exchange market, and we apply the formalism of the CTRW theory to financial tick data of the yen-dollar exchange rate. In Section 2, we introduce stochastic quantities included the volatility in the CTRW theory. In Section 3, the dynamical behavior of returns and volatilities for the yen-dollar exchange rate is mainly calculated numerically at the long-time limit. We conclude with some result and outlook in the final section.

## 2. VOLATILITIES IN THE CTRW

In this section, we will mainly focus on the CTRW theory in order to discuss the probability density function. Let $R(t)$ be the return defined by

$$R(t) = \ln [P(t+t_0)/P(t_0)] \quad (1)$$

where $P(t)$ is the price of the yen-dollar exchange rate at time $t$, and $t_0$ is an arbitrary time. The zero-mean return $Y(t)$ becomes

$$Y(t) = R(t) - <R(t)> \quad (2)$$

In our case, we introduce the pausing time density function and the transition probability. When $X_n = t_n - t_{n-1}$ and $\Delta Y_n = Y(t_n) - Y(t_{n-1})$, the pausing time density function and the jumping probability density function are, respectively, defined by

$$\psi(t)dt = \text{Prob}\{t < X_n < t + dt\} \quad (3)$$

and

$$j(l)dl = \text{Prob}\{l < \Delta Y_n < l + dl\}. \quad (4)$$

We let consider that the coupled probability density $\Psi(l, t)$ for jump $l$ and pausing-time $t$ is defined by

$$\Psi(l,t) = j(l)\psi(t), \quad (5)$$

where $j(l)$ is the jumping probability density dependent of the length between steps, and $\psi(t)$ is the pausing-time density. Summing over all $l$ with the periodic boundary condition, Eq. (1) becomes

$$\sum_l \Psi(l,t) = \sum_l j(l)\psi(t) = \psi(t) \quad (6)$$

The survival probability, $\Psi(t)$, which is stayed for the time $t$ after arriving at an arbitrary lattice point, can be expressed in terms of

$$\Psi(t) = 1 - \int_0^\infty dt \sum_l \Psi(l,t) = 1 - \int_0^\infty dt\, \psi(t) \quad (7)$$

Hence, after substituting Eq. (6) into Eq. (5), the Fourier-Laplace transform of the probability density function $P(k,u)$ is derived [16] as

$$P(k,u) = \sum_l \int_0^\infty dt P(l,t)\exp(-ikl - ut) = [1-\psi(u)]/u[1-\Psi(k,u)] \qquad (8)$$

where Eq. (8) is the so-called Montroll-Weiss equation. The Fourier-Laplace transform of Eq. (5), i.e. the generalized structure function, is presented by

$$\Psi(k,u) = j(k)\psi(u). \qquad (9)$$

In our model, the jumping probability density and the pausing time density in the high-frequency region is, respectively, given by the power-law such as

$$j(l) \sim l^{-\beta} \qquad (10)$$

and

$$\psi(t) \sim t^{-\alpha-1} \qquad (11)$$

Hence, using Eq. (8), the probability density function as $|l|\to\infty$ becomes

$$P(l,t) \sim |l|^{-\gamma} \qquad (12)$$

and the volatility, or the second moment of the return process, is obviously obtained as

$$<l^2(t)> \sim t^{\kappa} \qquad (13)$$

in the long time limit. The volatility can measure the efficiency of financial markets. This means that the larger the value of volatilities takes, the larger the mismatch between the selling and buying investers becomes in financial markets. It is generally well known from Eq. (13) in the CTRW theory that the second moment is proportional to $t^{\kappa}$ with the exponent $\kappa = 1$ for normal diffusion, $\kappa < 1$ for subdiffusion, and $\kappa > 1$ for superdiffusion. We can find the probability density function of random walkers by inversely Fourier-Laplace transforming of Eq. (8) when the generalized structure function is considered as one known quantity. We can also track and seek the distributed situation of random walkers from the first and second moments, if one cannot almost calculate numerically the probability density function.

## 3. NUMERICAL RESULTS

In order to analyze directly the scaling exponents of the statistical quantities from Eq. (10)-(13), we apply the formalism of the CTRW theory to the yen-dollar exchange rate. First of all, we introduce the yen-dollar exchange rate for two delivery dates of tick data: One analyzes minutely tick data for the period 2nd January - 20th February 2001, while the other analyzes ten-minutely tick data for the period 2nd January 2001 - 31th December 2001. We use two sets of tick data composed of the yen-dollar exchange rate in which two average times between ticks are about one and ten minutes. We numerically obtain from Fig. 1 that the jumping probability density is given by a power law $j(l) \sim l^{-\beta}$ with $\beta$ = (one minute) and 4.56 (ten minutes) from the continuous tick data of the price for the yen-dollar exchange rate. The pausing time density of the yen-dollar exchange rate is proportional to a power law $\psi(t) \sim t^{-\alpha-1}$ with the scaling exponent $\alpha = 2.18$ (one minute) and 2.60 (ten minutes), as shown in Fig. 2.

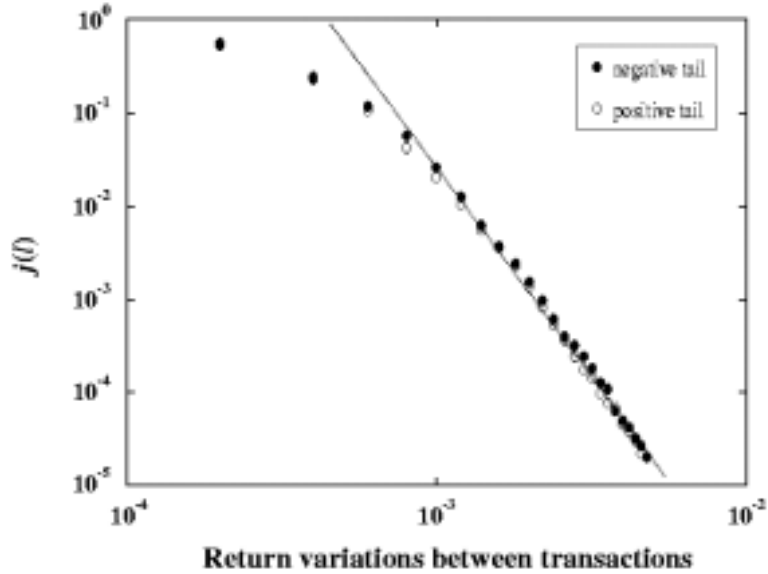

**Fig. 1** Jumping probability density of the yen-dollar exchange rate for one time step of ten minutes scales a power law $j(l) \sim l^{-\beta}$ with $\beta = 4.56$.

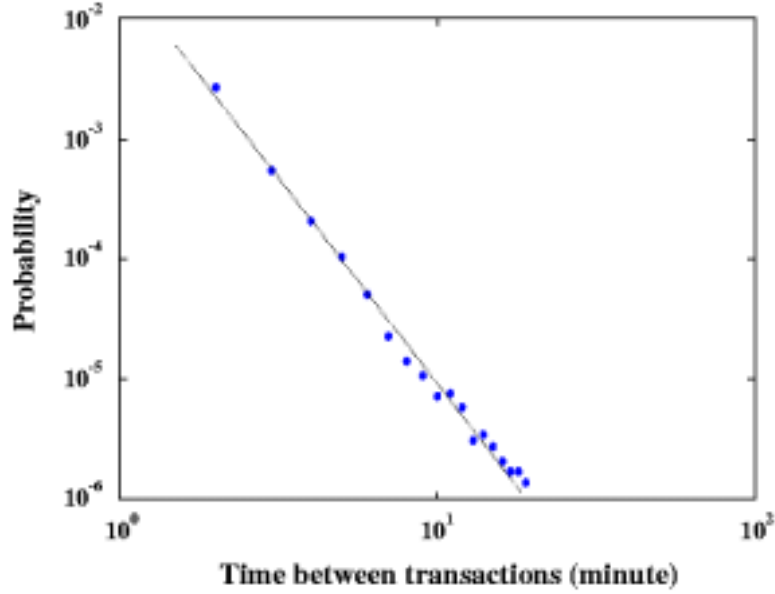

**Fig. 2** Pausing time density of the yen-dollar exchange rate for one time step of one minute showed a power law $\psi(t) \sim t^{-\alpha-1}$ with $\alpha = 2.60$.

For the probability density function, we obtained the scaling exponents $\gamma = 2.73$ and 2.07 (Fig. 3) for two cases of one time lag $t = 1$ minute and 10 minutes. Interestingly, the volatility is proportional anomalously to a power law with $\kappa = 0.96$ (one minute) and 0.86 (ten minutes) in Fig. 4. Our result is similar to that of IBM [9]. As a consequence, our case and IBM are subdiffusive process at the long time limit, while it behaves as the superdiffusive process for Gillette and as the diffusive process for dollar-deutsche mark exchange rate [9].

## 4. CONCLUSIONS

In conclusion, we have presented the dynamical behavior of the high-frequency tick data for yen-dollar exchange rate, based on the CTRW theory. We have numerically found the pausing time density, the jumping probability density, and the probability density function. Particularly, our case for the volatility is subdiffusive process at large time limits. It is in future expected that the detail description of the CTRW theory will be used to study the extension of financial analysis in the Korean and foreign financial markets.

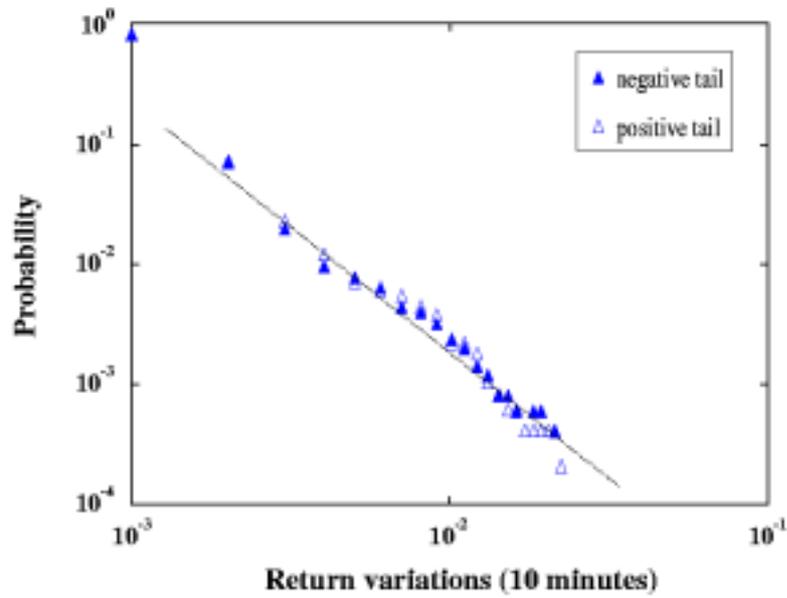

Fig. 3  Probability density function $P(l, t)$ of yen-dollar exchange rate for a time lag $t =$ ten minutes are proportional to a power law with the scaling exponent $\gamma = 2.07$.

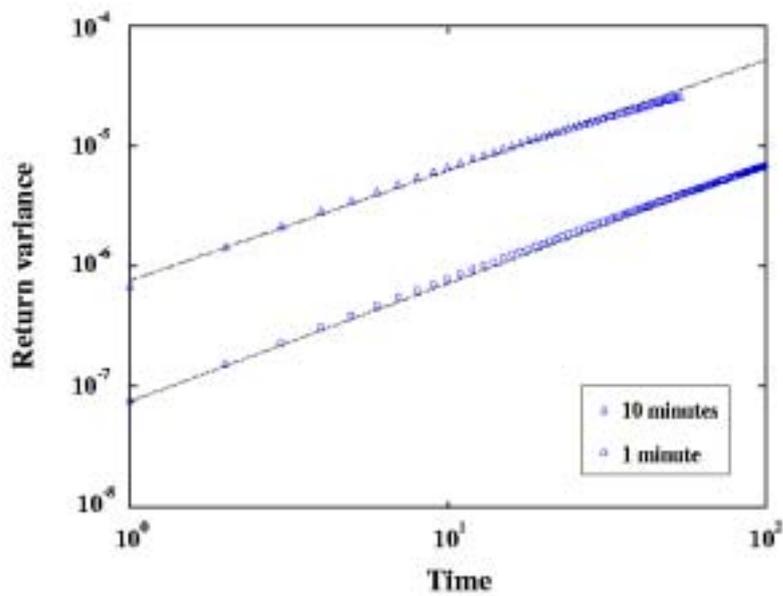

Fig. 4  Return variance is proportional to a power law with the scaling exponent $\kappa = 0.96$ (one minute) and 0.86 (ten minutes).

ACKNOWLEDGMENTS

This work was supported by Korea Research Foundation Grant(KRF-2004-002-B00026).